\documentclass{emulateapj}
\usepackage{apjfonts}
\usepackage{graphicx}

\shortauthors{Winn, Johnson, Albrecht, et al.~2009}
\shorttitle{Outlandish orbit of HAT-P-7b}

\begin{document}

% ------------------------------------------------------------------------
% New commands
%
\def\ltsima{$\; \buildrel < \over \sim \;$}
\def\lsim{\lower.5ex\hbox{\ltsima}}
\def\gtsima{$\; \buildrel > \over \sim \;$}
\def\gsim{\lower.5ex\hbox{\gtsima}}
                                                                                          
% -------------------------------------------------------------------------
%

\bibliographystyle{apj}

\title{ HAT-P-7: A Retrograde or Polar Orbit, and a Third Body }

\author{
Joshua N.\ Winn\altaffilmark{1},
John Asher Johnson\altaffilmark{2,3},
Simon Albrecht\altaffilmark{1},\\
Andrew W.\ Howard\altaffilmark{4,5},
Geoffrey W.\ Marcy\altaffilmark{4},
Ian J.\ Crossfield\altaffilmark{6},
Matthew J.\ Holman\altaffilmark{7}
}

\altaffiltext{1}{Department of Physics, and Kavli Institute for
  Astrophysics and Space Research, Massachusetts Institute of
  Technology, Cambridge, MA 02139}

\altaffiltext{2}{Institute for Astronomy, University of Hawaii,
  Honolulu, HI 96822}

\altaffiltext{3}{NSF Astronomy and Astrophysics Postdoctoral Fellow}

\altaffiltext{4}{Department of Astronomy, University of California,
  Mail Code 3411, Berkeley, CA 94720}

\altaffiltext{5}{Townes Postdoctoral Fellow, Space Sciences
  Laboratory, University of California, Berkeley, CA 94720}

\altaffiltext{6}{Department of Physics and Astronomy, University of
  California, Los Angeles, CA 90095}

\altaffiltext{7}{Harvard-Smithsonian Center for Astrophysics, 60
  Garden St., Cambridge, MA 02138}

\begin{abstract}

  We show that the exoplanet HAT-P-7b has an extremely tilted orbit,
  with a true angle of at least 86$^\circ$ with respect to its parent
  star's equatorial plane, and a strong possibility of retrograde
  motion. We also report evidence for an additional planet or
  companion star. The evidence for the unparalleled orbit and the
  third body is based on precise observations of the star's apparent
  radial velocity. The anomalous radial velocity due to rotation (the
  Rossiter-McLaughlin effect) was found to be a blueshift during the
  first half of the transit and a redshift during the second half, an
  inversion of the usual pattern, implying that the angle between the
  sky-projected orbital and stellar angular momentum vectors is
  $182\fdg 5\pm 9\fdg4$. The third body is implicated by excess
  radial-velocity variation of the host star over 2~yr. Some possible
  explanations for the tilted orbit are a close encounter with another
  planet, the Kozai effect, and resonant capture by an
  inward-migrating outer planet.

\end{abstract}


\section{Introduction}

In the Solar system, the planetary orbits are well-aligned and
prograde, revolving in the same direction as the rotation of the Sun.
This fact inspired the ``nebular hypothesis'' that the Sun and planets
formed from a single spinning disk (Laplace 1796). One might also
expect exoplanetary orbits to be well-aligned with their parent stars,
and indeed this is true of most systems for which it has been possible
to compare the directions of orbital motion and stellar rotation
(Fabrycky \& Winn 2009, Le Bouquin et al.~2009). However, there are at
least 3 exoplanets for which the orbit is tilted by a larger angle
than any of the planets in the Solar system: XO-3b (H\'ebrard et
al.~2008, Winn et al.~2009a), HD~80606b (Moutou et al.~2009, Pont et
al.~2009, Winn et al.~2009b), and WASP-14b (Johnson et al.~2009).

Still, all of those systems are consistent with prograde orbits, with
the largest minimum angle between the stellar-rotational and orbital
angular momentum vectors of about 37$^\circ$, for XO-3b (Winn et
al.~2009a). The reason why only the {\it minimum}\, angle is known is
that the evidence for misalignment is based on the eponymous effect of
Rossiter (1924) and McLaughlin (1924), an anomalous Doppler shift
observed during planetary transits that is sensitive only to the angle
between the {\it sky projections}\, of the two vectors. The true
spin-orbit angle may be larger, depending on the unknown inclination
angle of the stellar rotation axis with respect to the line of sight.

In this Letter we present evidence of a very large spin-orbit
misalignment for HAT-P-7b, a planet of mass 1.8~$M_{\rm Jup}$ and
radius 1.4~$R_{\rm Jup}$ in a 2.2-day orbit around an F6V star with
mass 1.5~$M_\odot$ and radius 1.8~$R_\odot$ (P\'al et al.~2008). We
find the angle between the sky-projected angular momentum vectors to
be $182\fdg 5\pm 9\fdg4$. Furthermore we show that the true angle
$\psi$ between those vectors is likely greater than $86^\circ$,
indicating that the orbit is either retrograde ($\psi > 90^\circ$) or
nearly polar ($\psi \approx 90^\circ$). We also present evidence for a
third body in the system, which may be an additional planet or a
companion star. We present spectroscopic data in \S~2, photometric
data in \S~3, a joint analysis of both types of data in \S~4, and a
discussion of the results in \S~5.

\section{Radial Velocities}
\label{sec:rv}

We observed HAT-P-7 with the High Resolution Spectrograph (HIRES) on
the Keck~I 10m telescope, and the High Dispersion Spectrograph (HDS)
on the Subaru~8m telescope. The planet's discoverers (P\'al et
al.~2008; hereafter, P08) obtained 8 HIRES spectra in 2007, to which
we add 9 spectra from 2009. All but one of the HIRES spectra were
acquired outside of transits. Of the 49 HDS spectra, 9 were obtained
on 2009~June~17 and 40 were obtained on 2009~July~1. The second of
these nights spanned a transit.

The instrument settings and observing procedures in both 2007 and 2009
were identical to those used by the California Planet Search (CPS;
Howard et al.~2009). We placed an iodine gas absorption cell into the
optical path, to calibrate the instrumental response and wavelength
scale. The radial velocity (RV) of each spectrum was measured with
respect to an iodine-free template spectrum, using the algorithm of
Butler et al.~(2006) with subsequent improvements. Measurement errors
were estimated from the scatter in the fits to individual spectral
segments spanning a few Angstroms. The RVs are given in Table~1.

\subsection{Evidence for a third body}
\label{subsec:oot}

\begin{figure*}[ht]
\begin{center}
  \leavevmode
\hbox{
  \epsfxsize=5in
  \epsffile{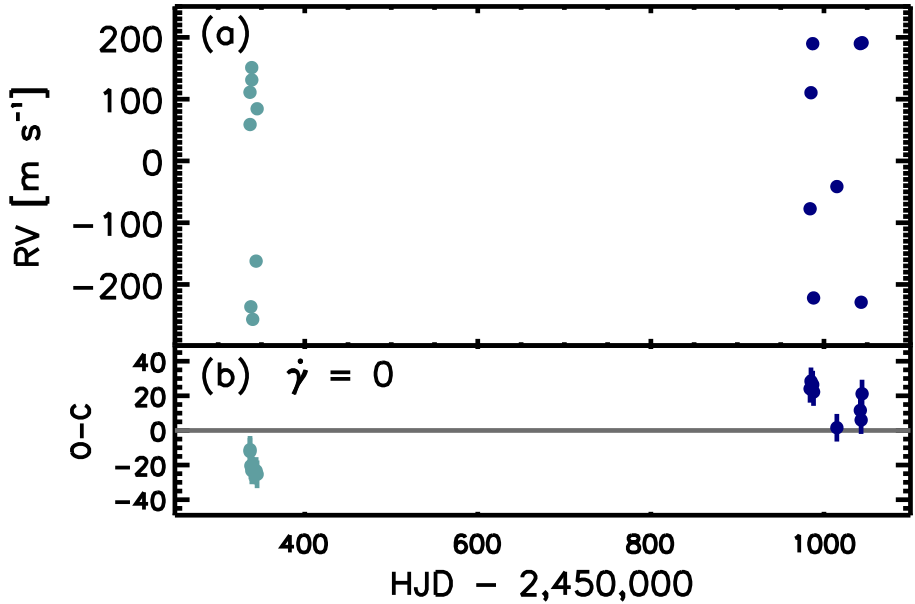}}
\end{center}
\caption{ \normalsize {\bf Long-term radial velocity variation of HAT-P-7.} (a)
  Measured RVs. (b) Residuals (observed~$-$~calculated) between the
  data and the best-fitting single-planet model. Light blue and dark
  blue points are HIRES data from 2007 and 2009,
  respectively. \label{fig:rv-time}}
\end{figure*}

Fig.~1 shows the RVs over the 2~yr span of the observations. Fig.~2
shows the RVs as a function of orbital phase, fitted with 2 different
models. The first model is a single Keplerian orbit, representing the
signal of the known planet. The second model has an additional
parameter $\dot{\gamma}$ representing an extra radial
acceleration. The second model gives a better fit to the data, with a
root-mean-squared (rms) residual of 7~m~s$^{-1}$ as compared to
21~m~s$^{-1}$ for the first model. The RVs from 2009 are
systematically redshifted by approximately 40~m~s~$^{-1}$ compared to
RVs from 2007, as evident from the residuals shown in Figs.~1(b) and
2(b). This shift is highly significant, as the CPS has demonstrated a
long-term stability of 2~m~s$^{-1}$ or better using HIRES and the same
reduction codes used here (Howard et al.~2009).

This RV trend is evidence for an additional companion. Given the
limited time coverage of our observations (two clusters of points
separated by 2~yr), the data are compatible with nearly any period
longer than a few months. A constant acceleration is the simplest
model that fits the excess RV variability, and under that assumption
we may give an order-of-magnitude relation relating $\dot{\gamma}$ to
some properties of the companion,
\begin{equation}
\frac{M_c \sin i_c}{a_c^2} \sim
\frac{\dot{\gamma}}{G} =
(0.121\pm 0.014)~M_{\rm Jup}~{\rm AU}^{-2},
\end{equation}
where $M_c$ is the companion mass, $i_c$ its orbital inclination
relative to the line of sight, $a_c$ its orbital distance, and the
numerical value is based on our model-fitting results (see \S~4).

\begin{figure*}[ht]
\begin{center}
  \leavevmode
\hbox{
  \epsfxsize=5in
  \epsffile{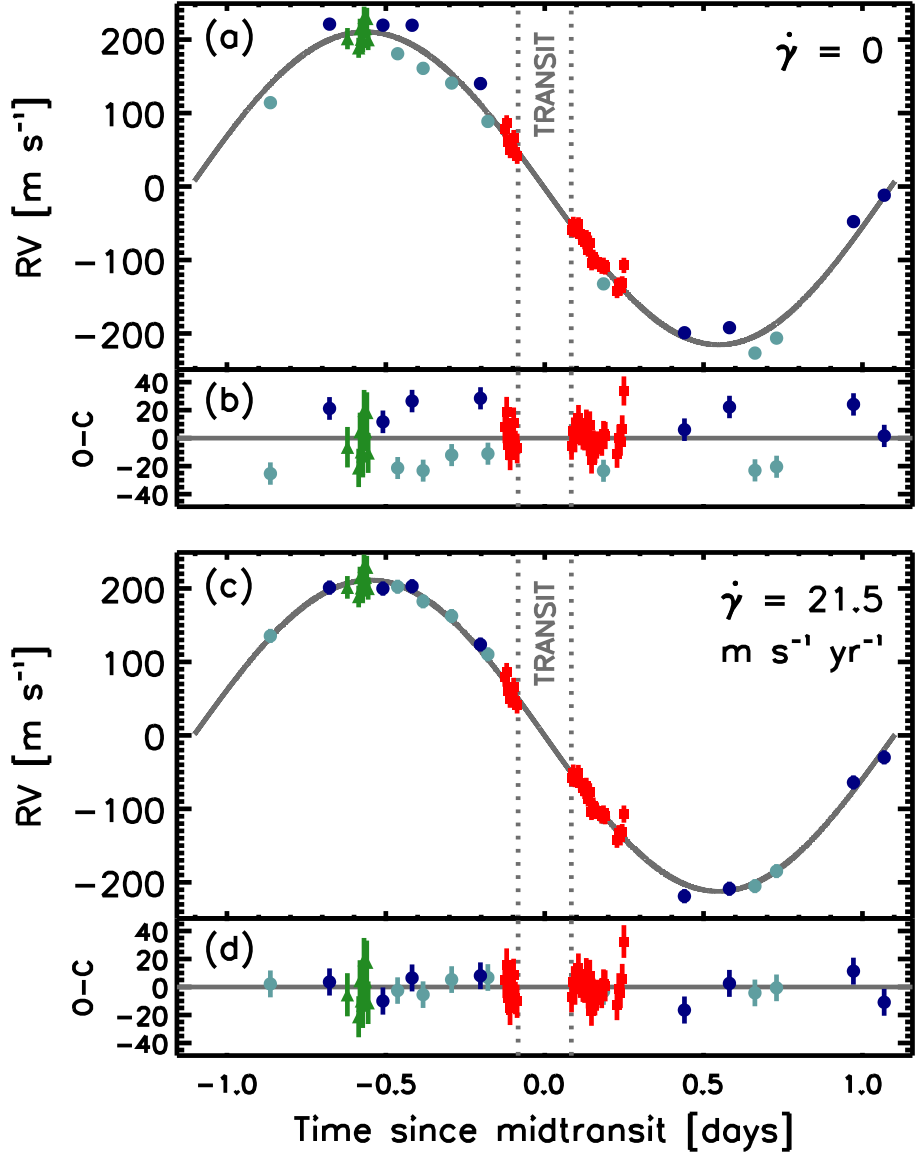}}
\end{center}
\caption{\normalsize {\bf Phased radial velocity variation of
    HAT-P-7.} (a) Assuming a single Keplerian orbit. (b)
  Residuals. (c) With an extra parameter $\dot{\gamma}$ representing a
  constant radial acceleration. (d) Residuals. The circles are HIRES
  data (light blue from 2007, dark blue from 2009), the green
  triangles are HDS data from 2009~June~17, and the red squares are
  HDS data from 2009~July~01.\label{fig:rv-phase}}
\end{figure*}

\subsection{Evidence for a spin-orbit misalignment}
\label{subsec:rm}

Fig.~3(a) shows the RV data spanning the transit, after subtracting
the orbital RV as computed with the best-fitting model including
$\dot{\gamma}$. We interpret the ``anomalous'' RV variation during the
transit as the Rossiter-McLaughlin (RM) effect, the asymmetry in the
spectral lines due to the partial eclipse of the rotating
photosphere. In the context of eclipsing binary stars, the RM effect
was predicted by Holt~(1893) and observed definitively by
Rossiter~(1924) and McLaughlin~(1924). For exoplanets, the RM effect
was first observed by Queloz et al.~(2000), and its use in assessing
spin-orbit alignment has been expounded by Ohta et al.~(2005) and
Gaudi \& Winn (2007).

A transiting planet in a well-aligned prograde orbit would first pass
in front of the blueshifted (approaching) half of the star, causing an
anomalous redshift of the observed starlight. Then, the planet would
cross to the redshifted (receding) half of the star, causing an
anomalous blueshift. In contrast, Fig.~3(a) shows a blueshift followed
by a redshift: an inversion of the effect just described. We may
conclude, even without any modeling, that the orbital ``north pole''
and the stellar ``north pole'' point in nearly opposite directions on
the sky.

\begin{figure*}[ht]
\begin{center}
  \leavevmode
\hbox{
  \epsfxsize=5in
  \epsffile{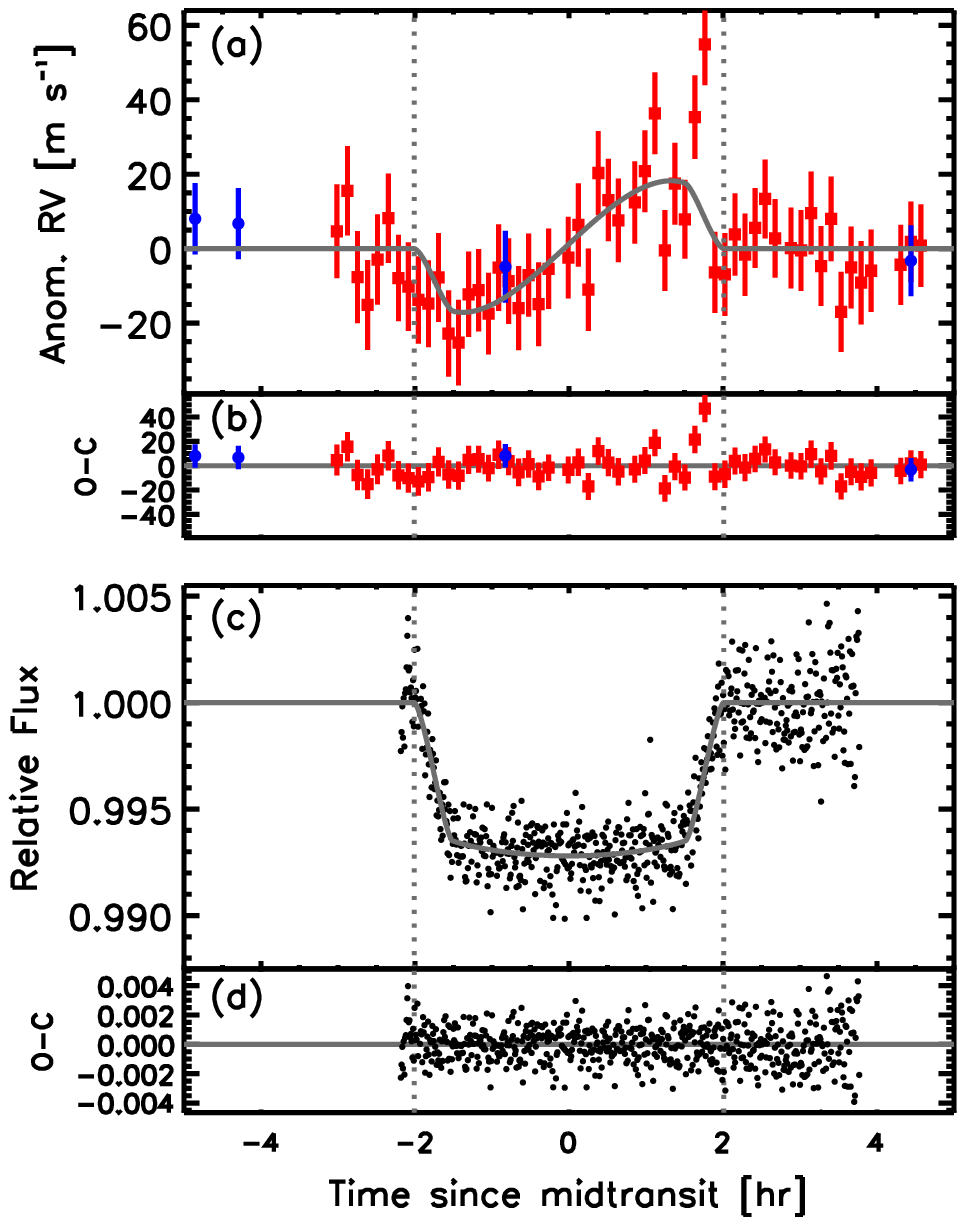}}
\end{center}
\caption{\normalsize {\bf The spectroscopic and photometric transit of
    HAT-P-7b.}  (a) The anomalous RV, defined as the output of the
  Doppler code minus the orbital RV. We observed a blueshift in the
  first half of the transit, and a redshift in the second half of the
  transit, demonstrating that the sky projections of the orbital and
  stellar angular momentum vectors point in opposite directions. (b)
  Residuals. Red squares are HDS data from 2009~July~1, and blue
  circles are HIRES data obtained on various nights in 2007 and
  2009. (c) The relative flux, observed in the Sloan $i$ band with the
  FLWO 1.2m telescope and Keplercam. (d) Residuals. In panels (a) and
  (b), the gray line shows the best fitting model.\label{fig:transit}}
\end{figure*}

\section{Photometry}
\label{sec:photometry}

For a quantitative analysis of the RM effect we wanted to model both
the photometric and spectroscopic transit signals. For this purpose we
supplemented the RV data with the most precise transit light curve
available to us, shown in Fig.~3(c). This light curve is based on
observations on UT~2008~Sep~22 in the Sloan $i$ bandpass, with the
Fred L.\ Whipple 1.2m telescope and Keplercam detector, under the
auspices of the Transit Light Curve project (Holman et al.~2006, Winn
et al.~2007).

Reduction of the CCD images involved standard procedures for bias
subtraction and flat-field division. Differential aperture photometry
was performed for HAT-P-7 and 7 comparison stars. No evidence was
found for time-correlated noise using the ``time-averaging'' method of
Pont et al.~(2006), as implemented by Winn et al.~(2009c). The data
shown in Fig.~3(c) were corrected for differential extinction as
explained in \S~4.

\section{Joint Analysis}
\label{sec:analysis}

We fitted a model to the photometric and RV data in order to derive
quantitative constraints on the angle $\lambda$ between the sky
projections of the orbital and stellar-rotational angular momentum
vectors. This angle is defined such that $\lambda=0^\circ$ when the
sky-projected vectors are parallel and $\lambda=180^\circ$ when they
are antiparallel. Our model for the RM effect was based on the
technique of Winn et al.~(2005): we simulated spectra exhibiting the
RM effect at various transit phases, and then measured the apparent RV
of the simulated spectra using the same Doppler code that is used on
actual data. This allowed us to relate the anomalous RV to the
parameters and positions of the star and planet.

The RV model was the sum of the Keplerian RV and the anomalous RV due
to the RM effect. The photometric model was based on the analytic
equation for the flux of a quadratically limb-darkened disk with a
circular obstruction (Mandel \& Agol 2002). As a compromise between
fixing the limb-darkening coefficients $u_1$ and $u_2$ at
theoretically calculated values, and giving them complete freedom, we
fixed $u_1-u_2$ at the tabulated value of 0.3846 (Claret 2004) and
allowed $u_1+u_2$ to be a free parameter. We also included a free
parameter for the coefficient of differential airmass extinction
between HAT-P-7 and the ensemble of comparison stars.

We determined the best values of the model parameters and their 68.3\%
confidence limits using a Markov Chain Monte Carlo algorithm, as
described in our previous works (see, e.g., Winn et al.~2009a). The
likelihood function was given by $\exp(-\chi^2/2)$ with
\begin{equation}
\chi^2 = \sum_{i=1}^{N_f} \left[ \frac{f_i({\mathrm{obs}}) - f_i({\mathrm{calc}})}{\sigma_{f,i}} \right]^2 +
         \sum_{i=1}^{N_v} \left[ \frac{v_i({\mathrm{obs}}) - v_i({\mathrm{calc}})}{\sigma_{v,i}} \right]^2,
\end{equation}
in a self-explanatory notation, with $\sigma_{f,i}$ chosen to be
0.00136, and $\sigma_{v,i}$ chosen to be the quadrature sum of the RV
measurement error and a ``stellar jitter'' term of
9.3~m~s$^{-1}$. These choices led to $\chi^2 = N_{\rm dof}$ for the
minimum-$\chi^2$ model. A Gaussian prior constraint was imposed upon
the orbital period based on the precise measurement of P08.

Table~2 gives the results for the model parameters. In particular, the
result for $\lambda$ is $182.5\pm 9.4$~deg, close to antiparallel, as
anticipated from the qualitative discussion of \S~2.

\section{Discussion}

Our finding for $\lambda$ is strongly suggestive of retrograde motion,
in which the orbital motion and stellar rotation are in opposite
directions. However, it must be remembered that $\lambda$ refers to
the angle between the {\it sky-projected}\, angular momentum
vectors. The true angle $\psi$ between the vectors is given by
\begin{equation}
\cos\psi = \cos i_\star \cos i + \sin i_\star \sin i \cos\lambda,
\end{equation}
where $i$ and $i_\star$ are the line-of-sight inclinations of the
orbital and stellar angular momentum vectors, respectively. Although
$i$ is known precisely from the transit data, $i_\star$ is unknown.

Supposing $i_\star$ to be drawn from an ``isotropic'' distribution
(uniform in $\cos i_\star$), the data demand that $\psi > 86\fdg 3$
with 99.73\% confidence. Thus, under this assumption, a retrograde
orbit is strongly favored, although a nearly-polar and barely-prograde
orbit cannot be ruled out.

In fact there is circumstantial evidence that $i_\star$ is small and
consequently the orbit of HAT-P-7b is nearly polar ($\psi \approx
90^{\circ}$). The star's projected rotation rate is unusually low for
such a hot star: $v\sin i_\star = 4.9_{-0.9}^{+1.2}$~km~s$^{-1}$ in
our model, or $3.8\pm 0.5$~km~s$^{-1}$ based on the line profile
analysis of P08; and $T_{\rm eff} = 6350\pm 80$~K according to P08. In
the SPOCS catalog of dwarf stars with well-determined spectroscopic
properties (Valenti \& Fischer 2005), only 2 of 37 stars with $T_{\rm
  eff} = 6350\pm 100$~K have $v\sin i_\star < 4.9$~km~s$^{-1}$.

Based on this catalog, the mean rotation rate $v$ for such hot stars
is about 15~km~s$^{-1}$. As an alternate approach to constraining
$\psi$, we assumed the rotation velocity $v$ is drawn from a Gaussian
distribution with mean 15~km~s$^{-1}$ and standard deviation
3~km~s$^{-1}$. The result is $\psi = 94.6_{-3.0}^{+5.5}$~deg with
68.3\% confidence, and $\psi > 86\fdg 1$ with 99.73\% confidence. This
analysis favors nearly-polar and retrograde orbits. However, one
wonders whether HAT-P-7 should be expected to have a ``typical''
rotation rate, given the existence of its short-period planet on a
bizarre orbit. Another caveat is that we found the scaled semimajor
axis $a/R_\star$ to be about 1$\sigma$ smaller than the finding of
P08, suggesting the star is somewhat larger and more evolved, which
would correspond to a slower expected rotation rate.

Determining $i_\star$ directly may be possible by measuring and
interpreting asteroseismological oscillations (Gizon \& Solanki 2003),
or photometric modulations produced by starspots (see, e.g., Henry \&
Winn 2008). By good fortune, HAT-P-7 is in the field of view of the
{\it Kepler}\, satellite, which is capable of precise long-term
photometry and may be able to accomplish these tasks (Borucki et
al.~2009).

The extraordinary orbit of HAT-P-7b presents an extreme case for
theories of planet formation and subsequent orbital evolution.
HAT-P-7b is a ``hot Jupiter'' and presumably migrated inward toward
the star after its formation. A prevailing migration theory involves
tidal interactions with the protoplanetary disk, but such interactions
would probably not perturb the initial coplanarity of the system, and
might even bring the system into closer alignment (Lubow \& Ogilvie
2001, Cresswell et al.~2007). More promising to explain HAT-P-7b are
scenarios involving few-body dynamics, as those scenarios are expected
to produce misalignments. In one scenario, close encounters between
planets throw a planet inward, where its orbit is ultimately shrunk
and circularized by tidal dissipation (Chatterjee et al.~2008,
Juri{\'c} \& Tremaine 2008). Another idea is based on the Kozai (1962)
effect, whereby the gravitational force from a distant body on a
highly inclined orbit strongly modulates an inner planet's orbital
eccentricity and inclination (Fabrycky \& Tremaine 2007). Recent
calculations showed that a combination of planet-planet scattering,
the Kozai effect, and tidal friction can lead to nearly-circular
retrograde orbits (Nagasawa et al.~2008). A third proposed scenario
involves an inward-migrating outer planet that captures an inner
planet into a mean motion resonance; if the inner planet avoids being
ejected or consumed by the star, it may be released on a
nearly-circular retrograde orbit (Yu \& Tremaine 2001).

The prospect of explaining HAT-P-7b's orbit through few-body dynamics
lends extra importance to measuring the mass and orbital parameters of
the third body. If it turns out to be a planet, then HAT-P-7b will be
only the second known case of a transiting planet accompanied by
another planet, the first being HAT-P-13b (Bakos et al.~2009). Such
systems are highly desirable because the unusually precise
measurements enabled by transit observations can be used to determine
whether the orbits are coplanar and give clues about the system's
dynamical history (Fabrycky 2009).

{\it Note added after submission.}---Narita et al.~(2009) report
independent evidence for a retrograde or polar orbit of HAT-P-7b,
based on Subaru/HDS spectra spanning the transit of 2008~May~30.

\acknowledgments We are grateful to Yasushi Suto and Ed Turner for
stimulating our interest in this subject; Norio Narita and his team
for sharing their data in advance of publication; Dan Fabrycky,
Guillaume H\'ebrard, Andr\'as P\'al, Darin Ragozzine, Scott Tremaine,
Bill Welsh, and the anonymous referee for helpful comments on the
manuscript; Akito Tajitsu, Tae-Soo Pyo, Mark Everett, Howard Isaacson,
and Zach Gazak for assistance with observing; G\'asp\'ar Bakos and
Joel Hartman for help obtaining telescope time; Eric Gaidos and Debra
Fischer for trading telescope time on short notice; and Hector
Balbontin for hospitality at Las Campanas Observatory where this
manuscript was written.

Some of the data presented herein were obtained at the W.M.~Keck
Observatory, which is operated as a scientific partnership among the
California Institute of Technology, the University of California, and
the National Aeronautics and Space Administration, and was made
possible by the generous financial support of the W.M.~Keck
Foundation. We extend special thanks to those of Hawaiian ancestry on
whose sacred mountain of Mauna Kea we are privileged to be
guests. Without their generous hospitality, the Keck observations
presented herein would not have been possible. J.A.J.\ gratefully
acknowledges support from the NSF Astronomy and Astrophysics
Postdoctoral Fellowship program (grant no.\ AST-0702821). S.A.\
acknowledges the support of the Netherlands Organisation for
Scientific Research (NWO). J.N.W.\ gratefully acknowledges support
from the NASA Origins program through awards NNX09AD36G and
NNX09AB33G, and from an MIT Class of 1942 Career Development
Professorship.

{\it Facilities:} \facility{Subaru (HDS)}, \facility{Keck:I (HIRES)},
\facility{FLWO:1.2m (Keplercam)}

\LongTables

\begin{deluxetable}{lccr}

\tabletypesize{\scriptsize}

\tablecaption{Relative Radial Velocity Measurements of HAT-P-7\label{tbl:rv}}
\tablewidth{0pt}
\tabletypesize{\scriptsize}

\tablehead{
\colhead{HJD} &
\colhead{RV [m~s$^{-1}$]} &
\colhead{Error [m~s$^{-1}$]} &
\colhead{Spec.\tablenotemark{a}}
}

\startdata
  $  2454336.73960$  &  $    111.08$  &  $   1.72$   &   1  \\
  $  2454336.85367$  &  $     58.89$  &  $   1.78$   &   1  \\
  $  2454337.76212$  &  $   -236.06$  &  $   1.70$   &   1  \\
  $  2454338.77440$  &  $    151.06$  &  $   1.54$   &   1  \\
  $  2454338.85456$  &  $    131.12$  &  $   1.57$   &   1  \\
  $  2454339.89886$  &  $   -256.42$  &  $   1.83$   &   1  \\
  $  2454343.83180$  &  $   -162.19$  &  $   1.75$   &   1  \\
  $  2454344.98805$  &  $     84.42$  &  $   2.15$   &   1  \\
  $  2454983.99020$  &  $    -77.44$  &  $   2.20$   &   1  \\
  $  2454985.02095$  &  $    110.41$  &  $   1.96$   &   1  \\
  $  2454987.01053$  &  $    189.80$  &  $   2.23$   &   1  \\
  $  2454988.00940$  &  $   -221.82$  &  $   2.28$   &   1  \\
  $  2455014.95322$  &  $    -41.51$  &  $   2.08$   &   1  \\
  $  2455016.05508$  &  $      4.10$  &  $   2.23$   &   1  \\
  $  2455042.03696$  &  $    189.91$  &  $   2.53$   &   1  \\
  $  2455042.98594$  &  $   -228.74$  &  $   2.40$   &   1  \\
  $  2455044.07369$  &  $    191.34$  &  $   2.55$   &   1  \\
  $  2455000.03550$  &  $    196.69$  &  $  12.26$   &   2  \\
  $  2455000.07080$  &  $    184.21$  &  $  11.29$   &   2  \\
  $  2455000.07507$  &  $    210.15$  &  $  11.25$   &   2  \\
  $  2455000.07933$  &  $    195.95$  &  $  11.94$   &   2  \\
  $  2455000.08361$  &  $    193.11$  &  $  13.30$   &   2  \\
  $  2455000.08787$  &  $    225.66$  &  $  12.45$   &   2  \\
  $  2455000.09214$  &  $    210.55$  &  $  11.67$   &   2  \\
  $  2455000.09641$  &  $    224.15$  &  $  12.23$   &   2  \\
  $  2455000.10068$  &  $    195.24$  &  $  11.97$   &   2  \\
  $  2455013.75919$  &  $     74.13$  &  $   8.47$   &   2  \\
  $  2455013.76487$  &  $     81.76$  &  $   7.53$   &   2  \\
  $  2455013.77030$  &  $     55.46$  &  $   8.11$   &   2  \\
  $  2455013.77573$  &  $     44.87$  &  $   7.60$   &   2  \\
  $  2455013.78114$  &  $     53.93$  &  $   7.71$   &   2  \\
  $  2455013.78703$  &  $     61.60$  &  $   7.50$   &   2  \\
  $  2455013.79245$  &  $     42.35$  &  $   6.83$   &   2  \\
  $  2455013.79787$  &  $     36.87$  &  $   7.31$   &   2  \\
  $  2455013.80330$  &  $     30.06$  &  $   7.10$   &   2  \\
  $  2455013.80871$  &  $     25.86$  &  $   6.95$   &   2  \\
  $  2455013.81412$  &  $     29.68$  &  $   7.44$   &   2  \\
  $  2455013.81955$  &  $     11.37$  &  $   6.86$   &   2  \\
  $  2455013.82496$  &  $      5.68$  &  $   6.69$   &   2  \\
  $  2455013.83039$  &  $     15.44$  &  $   6.63$   &   2  \\
  $  2455013.83580$  &  $     13.27$  &  $   5.85$   &   2  \\
  $  2455013.84121$  &  $      3.74$  &  $   5.69$   &   2  \\
  $  2455013.84664$  &  $     12.79$  &  $   6.91$   &   2  \\
  $  2455013.85206$  &  $      5.91$  &  $   6.67$   &   2  \\
  $  2455013.85748$  &  $     -4.64$  &  $   6.31$   &   2  \\
  $  2455013.86290$  &  $      0.95$  &  $   6.07$   &   2  \\
  $  2455013.86832$  &  $    -10.16$  &  $   6.26$   &   2  \\
  $  2455013.87374$  &  $     -3.95$  &  $   5.32$   &   2  \\
  $  2455013.88433$  &  $     -7.40$  &  $   5.70$   &   2  \\
  $  2455013.88976$  &  $     -1.91$  &  $   6.11$   &   2  \\
  $  2455013.89518$  &  $    -22.65$  &  $   5.91$   &   2  \\
  $  2455013.90060$  &  $      5.42$  &  $   6.32$   &   2  \\
  $  2455013.90603$  &  $     -5.14$  &  $   5.87$   &   2  \\
  $  2455013.91145$  &  $    -13.89$  &  $   6.19$   &   2  \\
  $  2455013.92034$  &  $    -14.44$  &  $   5.90$   &   2  \\
  $  2455013.92575$  &  $     -9.34$  &  $   5.80$   &   2  \\
  $  2455013.93118$  &  $      2.95$  &  $   5.79$   &   2  \\
  $  2455013.93660$  &  $    -37.18$  &  $   5.53$   &   2  \\
  $  2455013.94201$  &  $    -22.51$  &  $   5.80$   &   2  \\
  $  2455013.94744$  &  $    -35.31$  &  $   5.01$   &   2  \\
  $  2455013.95286$  &  $    -11.12$  &  $   6.17$   &   2  \\
  $  2455013.95829$  &  $      5.23$  &  $   5.75$   &   2  \\
  $  2455013.96372$  &  $    -59.30$  &  $   5.56$   &   2  \\
  $  2455013.96913$  &  $    -63.07$  &  $   6.04$   &   2  \\
  $  2455013.97456$  &  $    -55.50$  &  $   5.88$   &   2  \\
  $  2455013.98000$  &  $    -64.15$  &  $   5.92$   &   2  \\
  $  2455013.98542$  &  $    -60.13$  &  $   5.24$   &   2  \\
  $  2455013.99086$  &  $    -55.52$  &  $   4.96$   &   2  \\
  $  2455013.99627$  &  $    -69.24$  &  $   4.86$   &   2  \\
  $  2455014.00486$  &  $    -76.79$  &  $   5.51$   &   2  \\
  $  2455014.01029$  &  $    -80.53$  &  $   5.47$   &   2  \\
  $  2455014.01572$  &  $    -73.56$  &  $   5.96$   &   2  \\
  $  2455014.02114$  &  $    -90.86$  &  $   5.19$   &   2  \\
  $  2455014.02656$  &  $    -81.24$  &  $   6.44$   &   2  \\
  $  2455014.03199$  &  $   -109.25$  &  $   5.29$   &   2  \\
  $  2455014.03742$  &  $   -100.31$  &  $   5.77$   &   2  \\
  $  2455014.04285$  &  $   -107.39$  &  $   6.12$   &   2  \\
  $  2455014.04828$  &  $   -107.15$  &  $   5.91$   &   2  \\
  $  2455014.06427$  &  $   -114.12$  &  $   5.32$   &   2  \\
  $  2455014.06971$  &  $   -110.83$  &  $   5.39$   &   2  \\
  $  2455014.07514$  &  $   -114.67$  &  $   5.97$   &   2  \\
  $  2455014.11226$  &  $   -146.78$  &  $   5.93$   &   2  \\
  $  2455014.11769$  &  $   -137.88$  &  $   5.84$   &   2  \\
  $  2455014.12312$  &  $   -143.30$  &  $   6.07$   &   2  \\
  $  2455014.12855$  &  $   -136.48$  &  $   5.96$   &   2  \\
  $  2455014.13397$  &  $   -111.90$  &  $   7.02$   &   2
\enddata

\tablenotetext{a}{(1) Keck/HIRES, (2) Subaru/HDS.}

\tablecomments{The RV was measured relative to an arbitrary template
  spectrum specific to each spectrograph; only the differences among
  the RVs from a single spectrograph are significant. The uncertainty
  given in Column 3 is the internal error only and does not account
  for any possible ``stellar jitter.''}

\end{deluxetable}

\begin{deluxetable}{lc}
 
\tabletypesize{\footnotesize}
\tablecaption{Model Parameters for HAT-P-7b\label{tbl:params}}
\tablewidth{0pt}
 
\tablehead{
\colhead{Parameter} &
\colhead{Value}
}

\startdata
Orbital period, $P$~[d]                            & $2.2047304 \pm 0.0000024$    \\
Midtransit time~[HJD]                              & $2,454,731.67929 \pm 0.00043$  \\
Transit duration~(first to fourth contact)~[hr]    &  $4.006 \pm 0.064$             \\
Transit ingress or egress duration~[hr]            &  $0.474_{-0.093}^{+0.061}$       \\
\hline
Planet-to-star radius ratio, $R_p/R_\star$          & $0.0834_{-0.0021}^{+0.0012}$     \\
Orbital inclination, $i$~[deg]                     & $80.8_{-1.2}^{+2.8}$     \\
Scaled semimajor axis, $a/R_\star$                  &  $3.82_{-0.16}^{+0.39}$   \\
Transit impact parameter                           &  $0.618_{-0.149}^{+0.039}$ \\
\hline
Velocity semiamplitude, $K$~[m~s$^{-1}$]            & $211.8 \pm 2.6$ \\
Upper limit on eccentricity (99.73\% conf.)        & $ 0.039$ \\
$e\cos\omega$                                      & $-0.0019\pm 0.0077$ \\
$e\sin\omega$                                      & $ 0.0037\pm 0.0124$ \\
Velocity offset, Keck/HIRES [m~s$^{-1}$]            & $-51.2 \pm 3.6$ \\
Velocity offset, Subaru/HDS [m~s$^{-1}$]            & $-4.8 \pm 2.5$ \\
Constant radial acceleration $\dot{\gamma}$ [m~s$^{-1}$~yr$^{-1}$]  & $21.5\pm 2.6$ \\
\hline
Projected stellar rotation rate, $v \sin i_\star$~[km~s$^{-1}$]   &  $4.9_{-0.9}^{+1.2}$ \\
Projected spin-orbit angle, $\lambda$~[deg]         &  $182.5 \pm 9.4$
\enddata

\end{deluxetable}

\end{document}